\documentclass[conference]{IEEEtran}
\usepackage{cite} % Add the cite package to handle citations
\IEEEoverridecommandlockouts
\usepackage{hyperref}
\hypersetup{hidelinks}
\usepackage{amsmath, amssymb, amsfonts}
\usepackage{algorithmic}
\usepackage{graphicx}
\graphicspath{{images/}}
\usepackage{dblfloatfix} 
\usepackage{textcomp}
\usepackage{xcolor}
\usepackage{booktabs}
\usepackage{flushend}
\def\BibTeX{{\rm B\kern-.05em{\sc i\kern-.025em b}\kern-.08em
    T\kern-.1667em\lower.7ex\hbox{E}\kern-.125emX}}
\begin{document}

\bstctlcite{IEEEexample:BSTcontrol}

\title{
    Science Checker Reloaded: A Bidirectional Paradigm for Transparency  and Logical Reasoning
}

\author{
    \IEEEauthorblockN{
        \href{https://orcid.org/0000-0002-9420-8798}{Loïc Rakotoson},
    }
    \IEEEauthorblockA{
        \textit{Opscidia}\\
        Paris, France \\
        loic.rakotoson@opscidia.com
    }
    \and
    \IEEEauthorblockN{
        \href{https://orcid.org/0000-0002-2084-7618}{Sylvain Massip}
    }
    \IEEEauthorblockA{
        \textit{Opscidia}\\
        Paris, France \\
        sylvain.massip@opscidia.com
    }
    \and
    \IEEEauthorblockN{
        \href{https://orcid.org/0000-0003-0744-642X}{Fréjus A. A. Laleye}
    }
    \IEEEauthorblockA{
        \textit{Opscidia}\\
        Paris, France \\
        frejus.laleye@opscidia.com
    }
}

\maketitle

\begin{abstract}
Information retrieval is a rapidly evolving field. However it still faces significant limitations in the scientific and industrial vast amounts of information, such as semantic divergence and vocabulary gaps in sparse retrieval, low precision and lack of interpretability in semantic search, or hallucination and outdated information in generative models.
In this paper, we introduce a two-block approach to tackle these hurdles for long documents. The first block enhances language understanding in sparse retrieval by query expansion to retrieve relevant documents. The second block deepens the result by providing comprehensive and informative answers to the complex question using only the information spread in the long document, enabling bidirectional engagement. At various stages of the pipeline, intermediate results are presented to users to facilitate understanding of the system's reasoning.
We believe this bidirectional approach brings significant advancements in terms of transparency, logical thinking, and comprehensive understanding in the field of scientific information retrieval.    
\end{abstract}


\section{Introduction}
In recent years, advancements in Natural Language Processing have significantly reshaped the information retrieval landscape with the introduction of semantic vector search and large language models. The integration of dense retrieval into vector databases, which leverages low-dimensional contextual information, goes beyond sparse methods by tapping into nuanced semantic relationships, enriching the understanding of textual data.

While Best Match 25 (BM25) has limitations in addressing semantic divergence and vocabulary gaps, hindering its ability to capture the nuances of language effectively, it continues to dominate industry usage \cite{bm-25}. This prevalence can be ascribed to its simplicity and efficiency, coupled with the challenges posed by the lack of generalization, interpretability issues, and the black-box nature of Transformer-based dense methods. Despite the industry's ongoing reliance on BM25, dense retrieval methods have yet to reach maturity for widespread adoption, making BM25 the go-to method for its efficiency in industrial applications.

Moreover, the computational demands of Large Language Models (LLMs) in Retrieval Augmented Generation (Section \ref{secrag}) pose scalability concerns, making their seamless integration into real-world applications challenging. Additionally, these models inability to provide explanations for their answers and occasional tendency to hallucinate information \cite{halluci-1}\cite{halluci-2} create significant challenges, particularly in domains where accountability and reliability are paramount.
In addition, the cost of computing embeddings and the storage in a dedicated database required to save the embeddings of documents in a knowledge management context makes it difficult to adopt these methods and industrialize them in real-world applications \cite{green_ai_2020}\cite{musser2023cost}.

To address these limitations, our approach aims to go beyond technological innovation. We aim to address the fundamental issues of transparency and logical reasoning in answer generation and retrieval. By providing a clearer understanding of how the system arrives at its responses, we empower users with the ability to fact-check and exert control over the reasoning process. Aiming for this goal, our approach seeks to bridge the existing gaps and help in transparency and logical reasoning in scientific information retrieval.
The construction of the solution must diminish slow and heavy processes with relatively small improvements to simpler or lighter versions to obtain a better trade-off between performance and cost.
We are addressing in this work the context of open-domain query with scientific and technical documents, which are generally long and contain dense information.

This paper is organized as follows. Section \ref{secbackground} provides an overview of the background works in scientific information retrieval and retrieval augmented generation. Section \ref{secapproach} presents our approach, which consists of two blocks: document retrieval and answer generation. Section \ref{secevaluations} discusses the evaluations of our approach. Section \ref{secdiscussion} presents the discussion of our approach, concludes the paper and outlines future work.

\section{Background Works} \label{secbackground}
\subsection[Scientific Information Retrieval]{Scientific Information Retrieval}
Science based fact checking and question answering is a complex task due to the complexity of scientific language \cite{generating-scientific-claims} in comparison to general language. It is essential for combating the spread of misinformation and assisting researchers in knowledge discovery. Various methods have been proposed, including scientific claim generation, boolean question answering, and semi-automatic discovery of relevant expert opinions \cite{generating-scientific-claims}\cite{adefact-icwsm23}\cite{survey-scientific-fact-checking}. In \cite{science-checker}, authors proposed an extractive-boolean system with justification and contradiction resolution in yes/no/neutral questions related to biomedical studies.
However, the deployment and industrialization of these methods are not done since they do not integrate with any real usage \cite{usability-biomedical-qa}. These systems do not bring improvements in the selection of relevant documents, do not update their knowledge and are difficult to interact with. The introduction of chatbot-like systems with large language models in the industry has made it possible to interact with the system and to have a better understanding of the system's reasoning.

\subsection[RAG]{Retrieval Augmented Generation (RAG)} \label{secrag}
The usage of LLMs in fact checking has become increasingly important for accurate and credible information retrieval in complex knowledge-intensive tasks \cite{understanding-retrieval-augmentation}\cite{generate-rather-than-retrieve}, but it still faces challenges such as generating fictitious responses and hallucinations.

\begin{figure}[!b]
    \centerline{\includegraphics[width=0.5\textwidth]{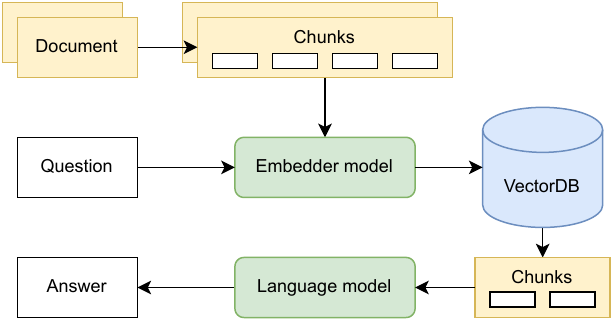}}
    \caption{Retrieval Augmented Generation Architecture}
    \label{fig:rag-archi}
\end{figure}

To overcome these challenges, researchers have proposed various approaches including the incorporation of Information Retrieval (IR) systems to provide external knowledge to LLMs. These approaches aim to improve the accuracy, credibility, and traceability of LLMs by verifying answers, correcting inaccuracies, and providing missing knowledge \cite{truth-o-meter}\cite{search-in-the-chain}. Additionally, domain-specific adaptations of LLMs have been explored to optimize performance in specialized domains \cite{reta-llm}. Furthermore, studies have examined the impact of retrieval augmentation on long-form question answering, including analysis of answer attribution and errors \cite{auto-claim-review}. Another perspective suggests replacing document retrievers with LLM generators for knowledge-intensive tasks, resulting in improved recall of acceptable answers \cite{improving-rag}.

\medskip
\subsubsection{Dense Retrieval}
To resolve the limitations of LLMs in knowledge update, vector databases have been proposed as a solution to the scalability and interpretability issues. These databases store dense representations of documents, enabling efficient retrieval of relevant documents \cite{generate-rather-than-retrieve}\cite{e-scan}\cite{vector-database-survey}.
The classic architecture of a RAG includes an ingestion module that divides each document into several chunks, with the document being stored in the form of embeddings of its chunks. A search module retrieves the embeddings of the most relevant chunks for a given question. A generation module takes the embeddings of the chunks as input and generates a response to the question (Figure \ref{fig:rag-archi}).

Aside from the computational costs of embeddings and inferences, which can be optimized through caching \cite{gpt-cache}, the storage of document embeddings in a dedicated database has a significant cost. In a knowledge management context with long documents and content-aware chunking, the storage space required for a document can be approximated by the following formula, without taking compression into account:
\begin{equation*}
    \text{Storage (bytes)} = \text{Chunks} \times \text{Embedding Size} + \text{Tokens}
\end{equation*}
The latest top 10 most performant embedding models in massive text embedding benchmark \cite{mteb-benchmark} have a maximum embedding size of 4096 dimensions and a minimum of 768 dimensions. The number of chunks in content-aware chunking is averaged at 4 to 12 sections per article. Which gives, for a small index of 10 million documents, a range of $0.1$ to $0.6$ TB of storage in a dense database. The same database in sparse representation will cost only $32$ GB.

In addition to their high cost, dense databases are less deployed in the industry due to their relatively low advantage compared to sparse databases. On top of this low efficiency, this immaturity is explained by the difficulty of adopting dense methods due to their lack of generalization, their low token-level performance, and especially the difficulty of interpreting the results \cite{dense-vs-sparse}\cite{simple-entity-centric} for end users.

\medskip
\subsubsection{Knowledge Graphs}
In the field of knowledge-intensive language tasks, innovative approaches have been developed to leverage external knowledge and enhance the capabilities of LLMs with knowledge graphs \cite{text-kgbench}. One such approach is the Knowledge Graph Induction framework \cite{kgi-framework}, which combines outputs from different models trained on tasks like slot filling, open domain question answering, dialogue, and fact-checking. This approach improves accuracy by cross-examining the outputs of various models, particularly enhancing dialogue using a question answering model. Another approach called Knowledge-Augmented language model Prompting (KAPING) \cite{knowledge-augmented} focuses on zero-shot knowledge graph question answering. This approach augments LLMs by retrieving relevant facts from a knowledge graph and incorporating them into the prompt without requiring additional model training. This method outperforms relevant zero-shot baselines by up to $48\%$ in average across multiple LLMs of varying sizes.
\newline
These approaches allow to increase the precision of the answers with easily verifiable knowledge and to make them more understandable for users.

\setcounter{figure}{2}
\begin{figure*}[b!]
    \centerline{\includegraphics[width=\textwidth]{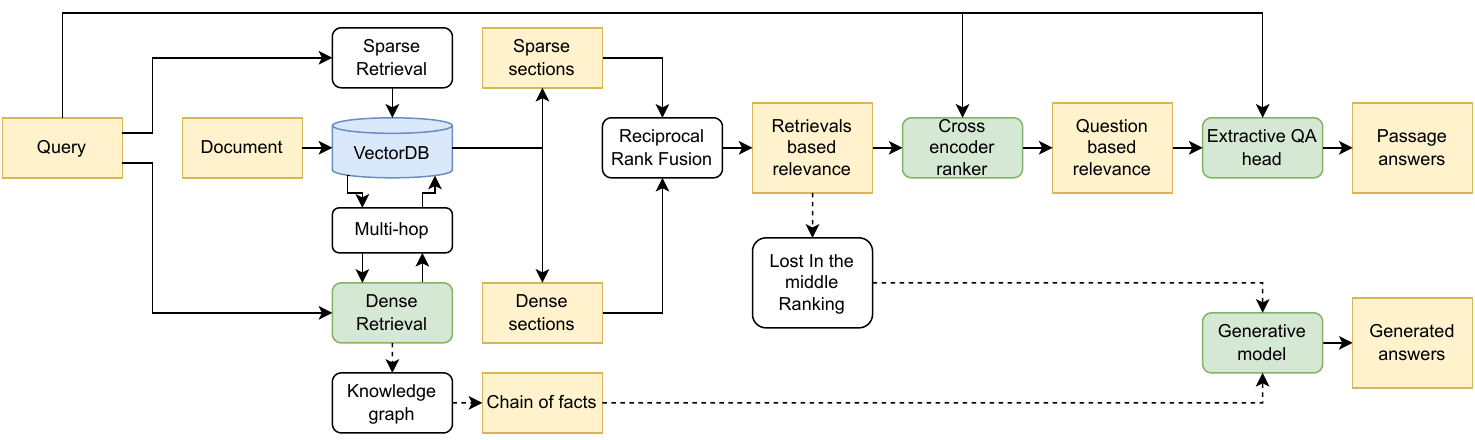}}
    \caption{
        Answer Generation with Iterative Deepening.
        Yellow: User access points.
    }
    \label{fig:answer-generation}
\end{figure*}

\section{Our Approach} \label{secapproach}
We aim to build an approach that can be industrialized, by eliminating the least efficient processes, allowing several pipeline stages to provide intermediate results to facilitate understanding of the system's reasoning, and enabling two-way interaction with the end user. The approach must follow the evolution of the user's funnel-like journey from information search to answer comprehension. The field of application is the retrieval of scientific and technical information, usually contained in long, dense documents.
\newline
We propose a two-block system. The first block focuses on retrieving relevant documents, while the second block iterates over each document to deepen the answer.

\setcounter{figure}{1}
\subsection{Document Retrieval}
\begin{figure}[ht!]
    \centerline{\includegraphics[width=0.4\textwidth]{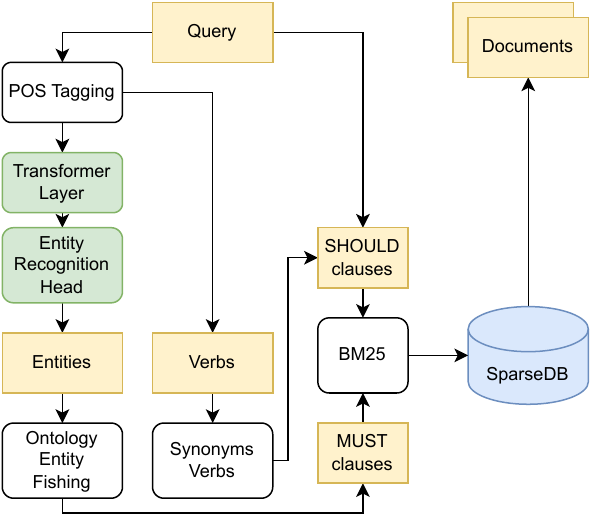}}
    \caption{
        Sparse Retrieval with Ontology-Oriented Query Expansion.
        Yellow: User access points.
    }
    \label{fig:sparse-retrieval}
\end{figure}
To retrieve relevant documents from a large collection based on a query, we apply an ontology-oriented sparse query expansion (Figure \ref{fig:sparse-retrieval}). We use the most efficient approach by supercharching BM25. To address the limitations of semantic divergence and vocabulary gaps, this approach leverages the semantic relationships between words and concepts to enrich the query with synonyms, hypernyms, and hyponyms.

The system extracts entities and performs expansion from the ontology. We use SciBERT \cite{scibert-19} as a model.
Each entity constitutes a MUST clause with its expansions as variations. Verbs are augmented with their Wordnet synonyms and will constitute SHOULD clauses. The database indexes the title, abstract, and sections of the article in different fields. The results are ranked by relevance and presented to the user.
Depending on the scope of knowledge management, the specificity of the ontology can be of different levels. We used Wikidata \cite{wikidata-taxonomy} as a generalist ontology, and some domain-specific ontologies like MeSH \cite{mesh-taxonomy}\cite{finance-taxonomy} or internal ontology.

\subsection{Answer Generation}
Once a relevant document has been identified by the user, they can query the long document more precisely to obtain an answer to their question. To achieve this, we implement a hybrid search system by creating an in-memory index of all the chunks of the document. The system includes a retrieval block consisting of sparse retrieval with a BM25 model and a dense retrieval in multihop \cite{multi-hop} with 3 iterations using multi-lingual semantic textual similarity sentence transformers \cite{multilingual-sentence-bert}.
The main output orders these results using a cross-encoder \cite{sentence-bert} and applies an extractive QA head to return the passages in the sections that answer the question.

Alternative outputs allow for a generative response.
The first uses the iterations of the multihop embedding retrieval to generate a sequence of logical reasoning from the selected chunks.
The second is placed at the output of the hybrid retrieval by scoring the sparse and dense results with a reciprocal rank fusion. Before feeding the generation model with this long context, a reranking in "lost in the middle" \cite{lost-in-the-middle} is performed to place the most relevant parts of the context at the beginning and end, using the primacy and recency biases of generative models.
The two outputs are under construction and their comparative evaluation between each other and with the primary output should validate the best approach for generating responses for end users according to different contexts of use and document nature.

In both cases, the generation model is used for its editorial and information synthesis functions from external sources of knowledge. As the model's internal knowledge should not be mobilized, the specialized language model is simpler, more efficient \cite{small-llm-1}\cite{small-llm-2}, and less prone to hallucination \cite{halluci-3}.

At various stages of the system, intermediate results are presented to the user to facilitate understanding of the system's reasoning. These access points are checkpoints for the user, who can thus verify the consistency of the results. Those access points are presented in yellow in Figures \ref{fig:sparse-retrieval} and \ref{fig:answer-generation}. They are the entry points for the user to interact with the system and to check the relevance of the sources and the reasoning of the system.

\section{Evaluations} \label{secevaluations}
The two blocks of our approach are systems that can be evaluated independently. To evaluate the usability of the whole, it is necessary to carry out user tests \cite{usability-biomedical-qa}, especially to evaluate the understanding of the response, the transparency of the system, and the ease of use. However, at the early stage of our work, we conducted quantitative performance evaluations of document retrieval. There is no benchmark sufficiently provided in long document retrieval on scientific data, however the Multilingual Long-Document Retrieval dataset built on Wikipedia and the associated work on M3-Embedding \cite{bge-m3} (Multi-Lingual, Multi-Functionality, Multi-Granularity Text Embeddings) are references to evaluate the performance of our document retrieval system. The dataset consists of text pairs of questions generated by GPT-3.5 (Generative Pre-trained Transformer) based on paragraphs sampled from lengthy articles in 13 languages, and the corresponding articles as the retrieval candidates. We focused on the English language.

We evaluated 3 versions of our system on an Elasticsearch base.
The first is the optimized use of the clauses of the base with different types of sub-fields of text, analyzers, and representations in n-grams. The clauses are boosted according to their degree of precision. As there is only one main textual field, the multi-match clauses are evaluated in most\_fields.
The second is the use of query expansion with an external knowledge graph. We used the Wikidata Ontology \cite{wikidata-taxonomy} to extend the entities detected in the query. These must respond in MUST clauses and boosted. The third is the same version with more flexible clauses on the recognized entities. We report the normalized Discounted Cumulative Gain (nDCG) among the top 10 results.

\begin{table}[ht]
    \centering
    \caption{
        \MakeLowercase{
            nDCG@10 of the different versions of our system and report of the last benchmark on the Multilingual Long-Document Retrieval dataset.
        }
    }
    \begin{tabular}{lcc}
        \toprule
        \textbf{Model} & \textbf{Max Length} & \textbf{nDCG@10} \\
        \toprule
        BM25                        & 8192  & 57.0 \\
        mDPR                        & 512   & 23.9 \\
        mContriveer                 & 512   & 28.7 \\
        mE5-large                   & 512   & 33.0 \\
        E5 Mistral 7B               & 8192  & 43.3 \\
        GPT-3 ada 002      & 8191  & 38.7 \\
        jina-embeddings-v2-base-en  & 8192  & 37.0 \\
        \midrule
        M3-Embedding              & & \\
        \midrule
        Dense                       & 8192 & 48.9 \\
        Sparse                      & 8192 & 62.1 \\
        Hybrid (Dense+Sparse)                & 8192 & \textbf{64.2} \\
        \midrule
        Ours                & & \\
        \midrule
        Optimized most\_fields      &   -  & 62.4 \\
        Must entity Wiki expansion  &   -  & 59.6 \\
        Should entity Wiki expansion&   -  & \textbf{64.8} \\
        \bottomrule
    \end{tabular}
\end{table}

While the first version gives 1,000 results for each query like the evaluation dataset, it turns out that the two versions with expansion give fewer search results. The absence of results is significant in the version with the expansion of entities executed in MUST clauses, with 18\% of queries without relevant documents. This is due to the strong precision constraint on the entities and results in a lower nDCG score than the version without expansion. By relaxing this constraint, with the SHOULD attribution clauses version, we improve the recall on the returned documents. This version, less complex and heavy than the hybrid version of M3-Embedding, gives performances that are roughly equivalent. In general, our approach, simpler and lighter, gives better performance than dense approaches based on more complex LLMs.
As with the second block, its results will need to be refined as the complete system is developed, with additional measures needed to capture the quality of the response, usability, and performance of the complete system in scientific and technical domains.

\section{Discussion} \label{secdiscussion}
Our approach is designed to address the limitations of existing systems in scientific information retrieval on long documents. It aims to provide a transparent and logical reasoning process, enabling users to fact-check and understand the system's reasoning. The system is designed to be industrialized, with a focus on eliminating the least efficient processes and being cost-aware. These early stage results are presented to highlight the principle of efficiency of accessible models in opposition to increasingly complex models, much more black-boxes.
The results of the document retrieval evaluation show that our approach, simpler and lighter, gives better performance than dense approaches based on heavier LLMs. This is a promising result for the industrialization of our approach. However, the evaluation of the complete system will require additional measures to capture the quality of the response, usability, and performance in scientific and technical domains.

In a knowledge management context on a specific domain, our retrieval system performs well. However, we observe the limitations of our current system due to the absence of a more general knowledge graph. In these specific cases, the use of the hybrid system proposed by M3-Embedding gives better results, despite these results being frozen by the model's weights. The development of top-level ontologies that allow the interoperability of more specific graphs, such as the WikiProject Ontology, is a method to have a system that improves with the advancement of knowledge. This is a development path for our system to have better knowledge coverage, in addition to regularly updated article databases.

Finally, as the goal of our system is to make exchanges between the pipeline and the user, we plan to include user tests in its development to evaluate the transparency of the system, ease of use, and understanding of the responses. These tests will validate the quality of our system and improve it. The ultimate goal is to propose a system that is easily industrializable and can be shaped according to deployment contexts. We conduct continuous tests on the system we are building during our research.

\section{Conclusion and Future Work}
In this paper, we presented a two-block approach for scientific information retrieval on long documents. Our approach combines sparse retrieval with ontology-oriented query expansion and hybrid retrieval with iterative deepening to provide comprehensive and informative answers to complex questions. We also designed our system to be transparent and logical, enabling oriented users evaluation with the system and understand its reasoning process. We evaluated our document retrieval block on the MLDR dataset and showed that our approach outperforms dense retrieval methods based on LLMs. The answer generation block is under evaluation. We plan to conduct user tests to evaluate the usability and quality of our complete system in scientific and technical domains. We also aim to improve our knowledge coverage by integrating top-level ontologies that allow the interoperability of more specific graphs. Moreover, we intend to explore the use of knowledge graphs and generative models to enhance the answer generation block and provide more credible and traceable responses. Finally, we hope to develop a system that is easily industrializable and adaptable to different deployment contexts.

\bibliographystyle{IEEEtran}
\bibliography{refs}

% Generated by IEEEtran.bst, version: 1.12 (2007/01/11)
\begin{thebibliography}{10}
\providecommand{\url}[1]{#1}
\csname url@samestyle\endcsname
\providecommand{\newblock}{\relax}
\providecommand{\bibinfo}[2]{#2}
\providecommand{\BIBentrySTDinterwordspacing}{\spaceskip=0pt\relax}
\providecommand{\BIBentryALTinterwordstretchfactor}{4}
\providecommand{\BIBentryALTinterwordspacing}{\spaceskip=\fontdimen2\font plus
\BIBentryALTinterwordstretchfactor\fontdimen3\font minus \fontdimen4\font\relax}
\providecommand{\BIBforeignlanguage}[2]{{%
\expandafter\ifx\csname l@#1\endcsname\relax
\typeout{** WARNING: IEEEtran.bst: No hyphenation pattern has been}%
\typeout{** loaded for the language `#1'. Using the pattern for}%
\typeout{** the default language instead.}%
\else
\language=\csname l@#1\endcsname
\fi
#2}}
\providecommand{\BIBdecl}{\relax}
\BIBdecl

\bibitem{bm-25}
M.~Luo, A.~Mitra, T.~Gokhale, and C.~Baral, ``Improving biomedical information retrieval with neural retrievers,'' in \emph{AAAI Conference on Artificial Intelligence}, vol.~36, no.~10, 2022, pp. 11\,038--11\,046.

\bibitem{halluci-1}
Z.~Ji \emph{et~al.}, ``Survey of hallucination in natural language generation,'' \emph{ACM Comput. Surv.}, vol.~55, no.~12, Mar 2023.

\bibitem{halluci-2}
L.~Huang \emph{et~al.}, ``A survey on hallucination in large language models: Principles, taxonomy, challenges, and open questions,'' \emph{ArXiv}, vol. abs/2311.05232, 2023.

\bibitem{green_ai_2020}
\BIBentryALTinterwordspacing
R.~Schwartz, J.~Dodge, N.~A. Smith, and O.~Etzioni, ``Green ai,'' \emph{Commun. ACM}, vol.~63, no.~12, p. 54–63, Nov 2020. [Online]. Available: \url{https://doi.org/10.1145/3381831}
\BIBentrySTDinterwordspacing

\bibitem{musser2023cost}
M.~Musser, ``A cost analysis of generative language models and influence operations,'' \emph{arXiv preprint arXiv:2308.03740}, 2023.

\bibitem{generating-scientific-claims}
D.~Wright \emph{et~al.}, ``Generating scientific claims for zero-shot scientific fact checking,'' in \emph{Proceedings of the 60th Annual Meeting of the Association for Computational Linguistics (Volume 1: Long Papers)}, 2022.

\bibitem{adefact-icwsm23}
E.~Altuncu \emph{et~al.}, ``aedfact: Scientific fact-checking made easier via semi-automatic discovery of relevant expert opinions,'' in \emph{Proceedings of the 17th International AAAI Conference on Web and Social Media}, 2023.

\bibitem{survey-scientific-fact-checking}
J.~Vladika and F.~Matthes, ``Scientific fact-checking: A survey of resources and approaches,'' in \emph{Findings of the Association for Computational Linguistics: ACL 2023}, 2023.

\bibitem{science-checker}
L.~Rakotoson, C.~Letaillieur, S.~Massip, and F.~A.~A. Laleye, ``Extractive-boolean question answering for scientific fact checking,'' in \emph{Proceedings of the 1st International Workshop on Multimedia AI against Disinformation}, ser. MAD '22.\hskip 1em plus 0.5em minus 0.4em\relax New York, NY, USA: Association for Computing Machinery, 2022, p. 27–34.

\bibitem{usability-biomedical-qa}
\BIBentryALTinterwordspacing
G.~Kell, I.~Marshall, B.~Wallace, and A.~Jaun, ``What would it take to get biomedical {QA} systems into practice?'' in \emph{Proceedings of the 3rd Workshop on Machine Reading for Question Answering}, A.~Fisch \emph{et~al.}, Eds.\hskip 1em plus 0.5em minus 0.4em\relax Punta Cana, Dominican Republic: Association for Computational Linguistics, Nov. 2021, pp. 28--41. [Online]. Available: \url{https://aclanthology.org/2021.mrqa-1.3}
\BIBentrySTDinterwordspacing

\bibitem{understanding-retrieval-augmentation}
H.-T. Chen, F.~Xu, S.~Arora, and E.~Choi, ``Understanding retrieval augmentation for long-form question answering,'' \emph{arXiv preprint arXiv:2310.12150}, 2023.

\bibitem{generate-rather-than-retrieve}
\BIBentryALTinterwordspacing
W.~Yu \emph{et~al.}, ``Generate rather than retrieve: Large language models are strong context generators,'' \emph{arXiv preprint arXiv:2209.10063}, 9 2022. [Online]. Available: \url{https://arxiv.org/pdf/2209.10063.pdf}
\BIBentrySTDinterwordspacing

\bibitem{truth-o-meter}
\BIBentryALTinterwordspacing
B.~Galitsky, ``Truth-o-meter: Collaborating with llm in fighting its hallucinations,'' \emph{Preprints}, 7 2023. [Online]. Available: \url{http://dx.doi.org/10.20944/preprints202307.1723.v1}
\BIBentrySTDinterwordspacing

\bibitem{search-in-the-chain}
S.~Xu, L.~Pang, H.~Shen, X.~Cheng, and T.-S. Chua, ``Search-in-the-chain: Towards accurate, credible and traceable large language models for knowledge-intensive tasks,'' \emph{arXiv preprint arXiv:2304.14732}, 2023.

\bibitem{reta-llm}
\BIBentryALTinterwordspacing
J.~Liu, J.~S. Jin, Z.~Wang, J.~Cheng, Z.~Dou, and J.-R. Wen, ``Reta-llm: A retrieval-augmented large language model toolkit,'' \emph{arXiv preprint arXiv:2306.05212}, 6 2023. [Online]. Available: \url{https://arxiv.org/pdf/2306.05212.pdf}
\BIBentrySTDinterwordspacing

\bibitem{auto-claim-review}
\BIBentryALTinterwordspacing
S.~Bhatia, J.~H. Lau, and T.~Baldwin, ``Automatic claim review for climate science via explanation generation,'' \emph{arXiv preprint arXiv:2107.14740}, 8 2021. [Online]. Available: \url{https://arxiv.org/pdf/2107.14740.pdf}
\BIBentrySTDinterwordspacing

\bibitem{improving-rag}
\BIBentryALTinterwordspacing
S.~Siriwardhana, R.~Weerasekera, E.~Wen, T.~Kaluarachchi, R.~Rana, and S.~Nanayakkara, ``Improving the domain adaptation of retrieval augmented generation ({RAG}) models for open domain question answering,'' \emph{Transactions of the Association for Computational Linguistics}, vol.~11, pp. 1--17, Jan 2023. [Online]. Available: \url{https://aclanthology.org/2023.tacl-1.1}
\BIBentrySTDinterwordspacing

\bibitem{e-scan}
V.~Sanca and A.~Ailamaki, ``E-scan: Consuming contextual data with model plugins,'' in \emph{Joint Workshops at 49th International Conference on Very Large Data Bases (VLDBW’23)}, 2023.

\bibitem{vector-database-survey}
Y.~Han, C.~Liu, and P.~Wang, ``A comprehensive survey on vector database: Storage and retrieval technique, challenge,'' \emph{arXiv preprint arXiv:2310.11703}, 2023.

\bibitem{gpt-cache}
\BIBentryALTinterwordspacing
F.~Bang, ``{GPTC}ache: An open-source semantic cache for {LLM} applications enabling faster answers and cost savings,'' in \emph{Proceedings of the 3rd Workshop for Natural Language Processing Open Source Software (NLP-OSS 2023)}, L.~Tan, D.~Milajevs, G.~Chauhan, J.~Gwinnup, and E.~Rippeth, Eds.\hskip 1em plus 0.5em minus 0.4em\relax Singapore: Association for Computational Linguistics, Dec. 2023, pp. 212--218. [Online]. Available: \url{https://aclanthology.org/2023.nlposs-1.24}
\BIBentrySTDinterwordspacing

\bibitem{mteb-benchmark}
N.~Muennighoff, N.~Tazi, L.~Magne, and N.~Reimers, ``Mteb: Massive text embedding benchmark,'' \emph{arXiv preprint arXiv:2210.07316}, 2022.

\bibitem{dense-vs-sparse}
N.~Arabzadeh, X.~Yan, and C.~L.~A. Clarke, ``Predicting efficiency/effectiveness trade-offs for dense vs. sparse retrieval strategy selection,'' \emph{arXiv preprint arXiv:2109.10739}, 2021.

\bibitem{simple-entity-centric}
\BIBentryALTinterwordspacing
C.~Sciavolino, Z.~Zhong, J.~Lee, and D.~Chen, ``Simple entity-centric questions challenge dense retrievers,'' in \emph{Proceedings of the 2021 Conference on Empirical Methods in Natural Language Processing}, M.-F. Moens, X.~Huang, L.~Specia, and S.~W.-t. Yih, Eds.\hskip 1em plus 0.5em minus 0.4em\relax Online and Punta Cana, Dominican Republic: Association for Computational Linguistics, Nov. 2021, pp. 6138--6148. [Online]. Available: \url{https://aclanthology.org/2021.emnlp-main.496}
\BIBentrySTDinterwordspacing

\bibitem{text-kgbench}
N.~Mihindukulasooriya, S.~Tiwari, C.~F. Enguix, and K.~Lata, ``Text2kgbench: A benchmark for ontology-driven knowledge graph generation from text,'' \emph{arXiv preprint arXiv:2308.02357}, 2023.

\bibitem{kgi-framework}
\BIBentryALTinterwordspacing
M.~F.~M. Chowdhury, M.~Glass, G.~Rossiello, A.~Gliozzo, and N.~Mihindukulasooriya, ``Kgi: An integrated framework for knowledge intensive language tasks,'' \emph{arXiv preprint arXiv:2204.03985}, 9 2022. [Online]. Available: \url{https://arxiv.org/pdf/2204.03985.pdf}
\BIBentrySTDinterwordspacing

\bibitem{knowledge-augmented}
\BIBentryALTinterwordspacing
J.~Baek, A.~F. Aji, and A.~Saffari, ``Knowledge-augmented language model prompting for zero-shot knowledge graph question answering,'' in \emph{Proceedings of the First Workshop on Matching From Unstructured and Structured Data (MATCHING 2023)}.\hskip 1em plus 0.5em minus 0.4em\relax Toronto, ON, Canada: Association for Computational Linguistics, Jul. 2023, pp. 70--98. [Online]. Available: \url{http://dx.doi.org/10.18653/v1/2023.matching-1.7}
\BIBentrySTDinterwordspacing

\bibitem{scibert-19}
\BIBentryALTinterwordspacing
I.~Beltagy, K.~Lo, and A.~Cohan, ``{S}ci{BERT}: A pretrained language model for scientific text,'' in \emph{Proceedings of the 2019 Conference on Empirical Methods in Natural Language Processing and the 9th International Joint Conference on Natural Language Processing (EMNLP-IJCNLP)}, K.~Inui, J.~Jiang, V.~Ng, and X.~Wan, Eds.\hskip 1em plus 0.5em minus 0.4em\relax Hong Kong, China: Association for Computational Linguistics, Nov. 2019, pp. 3615--3620. [Online]. Available: \url{https://aclanthology.org/D19-1371}
\BIBentrySTDinterwordspacing

\bibitem{wikidata-taxonomy}
F.~Brasileiro, J.~a. P.~A. Almeida, V.~A. Carvalho, and G.~Guizzardi, ``Applying a multi-level modeling theory to assess taxonomic hierarchies in wikidata,'' in \emph{Proceedings of the 25th International Conference Companion on World Wide Web}, ser. WWW '16 Companion.\hskip 1em plus 0.5em minus 0.4em\relax Republic and Canton of Geneva, CHE: International World Wide Web Conferences Steering Committee, 2016, p. 975–980.

\bibitem{mesh-taxonomy}
F.~B. Rogers, ``\BIBforeignlanguage{en}{Medical subject headings},'' \emph{\BIBforeignlanguage{en}{Bull. Med. Libr. Assoc.}}, vol.~51, pp. 114--116, Jan. 1963.

\bibitem{finance-taxonomy}
D.~Altinok, ``An ontology-based dialogue management system for banking and finance dialogue systems,'' \emph{arXiv preprint arXiv:1804.04838}, 2018.

\bibitem{multi-hop}
W.~Xiong \emph{et~al.}, ``Answering complex open-domain questions with multi-hop dense retrieval,'' \emph{arXiv preprint arXiv:2009.12756}, 2021.

\bibitem{multilingual-sentence-bert}
\BIBentryALTinterwordspacing
N.~Reimers and I.~Gurevych, ``Making monolingual sentence embeddings multilingual using knowledge distillation,'' in \emph{Proceedings of the 2020 Conference on Empirical Methods in Natural Language Processing}.\hskip 1em plus 0.5em minus 0.4em\relax Association for Computational Linguistics, 11 2020. [Online]. Available: \url{https://arxiv.org/abs/2004.09813}
\BIBentrySTDinterwordspacing

\bibitem{sentence-bert}
\BIBentryALTinterwordspacing
------, ``Sentence-bert: Sentence embeddings using siamese bert-networks,'' in \emph{Proceedings of the 2019 Conference on Empirical Methods in Natural Language Processing}.\hskip 1em plus 0.5em minus 0.4em\relax Association for Computational Linguistics, 11 2019. [Online]. Available: \url{https://arxiv.org/abs/1908.10084}
\BIBentrySTDinterwordspacing

\bibitem{lost-in-the-middle}
N.~F. Liu \emph{et~al.}, ``Lost in the middle: How language models use long contexts,'' \emph{arXiv preprint arXiv:2307.03172}, 2023.

\bibitem{small-llm-1}
W.~Shen \emph{et~al.}, ``Small llms are weak tool learners: A multi-llm agent,'' \emph{arXiv preprint arXiv:2401.07324}, 2024.

\bibitem{small-llm-2}
G.~Juneja, S.~Dutta, S.~Chakrabarti, S.~Manchanda, and T.~Chakraborty, ``Small language models fine-tuned to coordinate larger language models improve complex reasoning,'' \emph{arXiv preprint arXiv:2310.18338}, 2023.

\bibitem{halluci-3}
S.~Verma, K.~Tran, Y.~Ali, and G.~Min, ``Reducing llm hallucinations using epistemic neural networks,'' \emph{arXiv preprint arXiv:2312.15576}, 2023.

\bibitem{bge-m3}
J.~Chen, S.~Xiao, P.~Zhang, K.~Luo, D.~Lian, and Z.~Liu, ``Bge m3-embedding: Multi-lingual, multi-functionality, multi-granularity text embeddings through self-knowledge distillation,'' \emph{arXiv preprint arXiv:2402.03216}, 2024.

\end{thebibliography}
\end{document}